\documentclass[letter]{aa}     

\usepackage{CJKutf8}  
\usepackage{geometry}
\usepackage{graphicx}
\usepackage{txfonts}
\usepackage{amsmath}
\usepackage{amssymb}
\usepackage{aas_macros}
\usepackage{booktabs}
\usepackage{graphicx}
\usepackage{hyperref}
\usepackage{listings}
\usepackage{multirow}
\usepackage{extarrows}
\usepackage{bm}
\usepackage{wasysym}

\newcommand{\UTC}{\mathrm{UTC}}
\newcommand{\TCL}{\mathrm{TCL}}

\newcommand{\M}{\mathrm{M}}

\newcommand{\LL}{L_\mathrm{L}}

\newcommand{\tao}{time aligned orbit}

\begin{document}
\begin{CJK*}{UTF8}{gbsn}  
   \title{Two birds with one stone: simultaneous realization of both Lunar Coordinate Time and lunar geoid time by a single orbital clock}



   \author{  Tian-Ning Yang  \inst{1,2}
   \and Ren-Fang Geng  \inst{3}
   \and Jing Zhang \inst{1}
   \and Chong Yang \inst{1,2}
   \and Yong Huang \inst{3,4}
        \and Yi Xie \inst{1}
        }

   \institute{Purple Mountain Observatory, Chinese Academy of Sciences, Nanjing 210023, China\\
   \email{yixie@pmo.ac.cn}
            \and School of Astronomy and Space Science, University of Science and Technology of China, Hefei 230026, China 
            \and Shanghai Astronomical Observatory, Chinese Academy of Sciences, Shanghai 200030, China
            \and School of Astronomy and Space Science, University of Chinese Academy of Sciences, Beijing 100049, China\\}

   \date{Received XX, 2026}

 
  \abstract
  {Among options for definition of the lunar reference time, 
  the option taking Lunar Coordinate Time (O1) has its simplicity but cannot be realized by any clock without steering, 
  while another option adopting the lunar geoid (selenoid) proper time (O2) has its convenience for users on the lunar surface but would bring a new scaling of spatial coordinates and mass parameter of the Moon. }
  {We propose a ``time aligned orbit'' that the readings of an ideal clock in this orbit could equal to the selenoid proper time in O2 and these readings could be converted to Lunar Coordinate Time in O1 by a known linear transformation. } 
   {We show that there exist the time aligned orbit around the Moon with its semi-major axis of about 1.5 lunar radius slightly depending on its inclination.
   We conduct a set of numerical simulations to assess to what extent a clock on these orbits could realize O2 in a more realistic lunar environment.}
  {We find that the proper time in our simulations would desynchronize from the selenoid proper time up to 190 ns after a year with a frequency offset of $6\times10^{-15}$,
  which is solely 3.75\% of the frequency difference in O2 caused by the lunar surface topography.
  These numbers might be further reduced to $13$ ns and $4\times 10^{-16}$, if we could account for the deviation of the mean orbits in our simulations from the nominal ones.}
  { One might simultaneously realize O1 and O2 by deployment of a single clock in the time aligned orbit. This approach also has its scalability for other terrestrial planets beyond the Earth-Moon system. }

\titlerunning{Realization of TCL and selenoid time by an orbital clock}
   \maketitle

\nolinenumbers


\section{Introduction}

Definition of the lunar reference time (LRT) \citep{IAU2024ResIII} has drawn much attention recently.
It would satisfy the following criteria that (but may not limited to) \citep{Bourgoin2025arXiv2507.21597}
\begin{itemize}
    \item C1: LRT should be defined from Lunar Coordinate Time (TCL) \citep{IAU2024ResII}.
    \item C2: It should have a physical and available realization.
    \item C3: It should have a clear relationship with the Coordinated Universal Time (UTC).
    \item C4: It should be scalable to space beyond the Earth-Moon system.
\end{itemize}
The first criterion leads to three options that \citep{Bourgoin2025arXiv2507.21597}
\begin{itemize}
    \item O1: LRT is exactly the same as TCL.
    \item O2: LRT has the same average rate of the proper time of a clock on a given lunar geoid, i.e., selenoid.
    \item O3: LRT deviates form the Terrestrial Time (TT) solely by periodic variations. 
\end{itemize}

With its obvious advantage of simplicity, O1 has it distinct disadvantage that none of ideal clocks can realize TCL without steering in principle.
Although O2 and O3 could provide convenience for users on the lunar surface and those using Earth navigation satellite signals respectively, their scalings of TCL have to imply the same scalings of spatial coordinates and mass parameter of the Moon, causing possible confusion.
Moreover, it is still very challenging for successfully deploying and maintaining clocks on the surface of the Moon for O2 with the same logic as for realization of TT on the Earth.

With an attempt to reconcile these pros and cons, we propose a way that might simultaneously realize O1 and O2.
We show that there exists an orbit around the Moon
so that the readings of an ideal clock in this orbit equal to the readings of the proper time of an ideal clock on a given selenoid (O2).
Meanwhile, one might find TCL (O1) by scaling the readings of such an orbital clock with a factor related to the potential of the selenoid.
We call this orbit as the ``time aligned orbit''.

In Sec.~\ref{sec:theory}, we explain the underlying reasons for the existence of the \tao.
We present the properties of the \tao\ in a more realistic lunar environment by numerical simulation in Sec.~\ref{sec:sim}.
We conclude this work and discuss its scalability for other planets in Sec.~\ref{sec:con}.

\section{Theory}
\label{sec:theory}
In this section, we would explain the reason why a \tao\ exists around the Moon and extend this conception into a more generic case.
Under the Lunar Celestial Reference System (LCRS) \citep{IAU2024ResII}, its coordinate time TCL and the proper time $\tau$ of a clock in the vicinity of the Moon satisfy the following relation that \citep{Bourgoin2025arXiv2507.21597}
\begin{equation}
\label{eq:tau2TCL}
	\mathrm{TCL}-\mathrm{TCL}_0= \tau-\tau_0 + \frac{1}{c^2} \int_{\tau_0}^{\tau}\left[U_\M(\bm{Y})+\frac{\bm{\dot{Y}}^{2}}{2}\right]\mathrm{~d}\tau + \mathcal{O}(c^{-4}),
\end{equation}
where $\tau_0$ is the initial reading of the clock, $\TCL_0$ is the TCL moment corresponding to $\tau_0$, 
$\bm{Y}$ and $\dot{\bm{Y}}$ are the position and velocity vectors of the clock,
and $U_\M(\bm{Y})$ represents the gravitational potential at the clock from the Moon with the ignorance of effects from all of the other bodies in the Solar System. 

Considering the proper time $\tau_\mathrm{s}$ of an ideal clock on the equator of a given seleonid with its specified potential $W_{\M0}$, Eq.~\eqref{eq:tau2TCL} leads to \citep{Nelson2011Metrologia48.S171}
\begin{equation}
\label{eq:taus2TCL}
	\TCL = \left(1+\LL\right)(\tau_\mathrm{s}-\tau_\mathrm{s0})+\TCL_{\mathrm{s0}}+\mathcal{O}(c^{-4})
\end{equation}
with
\begin{equation}
\label{eq:LL}
  \LL\equiv c^{-2}W_{\M0} \approx   \frac{GM_{\mathrm{M}}}{c^{2}R_{\M}} \left(1+\frac{1}{2} J_{2}^{\M}+\frac{1}{2}\eta_{\M}\right),
\end{equation}
where $M_{\mathrm{M}}$, $R_{\M}$ and $J_{2}^{\M}$ are the mass, radius and dynamical form factor of the Moon, and we neglect the higher-order spherical harmonic components of $U_\M$.
The ratio of $\eta_{\M}$ is defined as
\begin{equation}
\eta_{\M} = \frac{ \bm{\dot{Y}}_{\M}^{2} }{ \mathcal{V}_{\M}^2 }
\end{equation}
where $\bm{\dot{Y}}_{\M}$ is the rotational velocity at the location of the clock on the surface of the Moon,
and $\mathcal{V}_{\M}=\sqrt{GM_{\M}/R_{\M}}$ is the first cosmic velocity of the Moon.

For the proper time $\tau_\mathrm{p}$ of an ideal clock in a mean circular orbit ($\bar{e}_\mathrm{p}=0$) around the Moon, Eq.~\eqref{eq:tau2TCL} gives \citep{Kouba2004GPSSol8.170,Formichella2021GPSSol25.56}
\begin{eqnarray}
\label{eq:taup2TCL}
	 \mathrm{TCL} = \left(1+L_\mathrm{P}\right) (\tau_\mathrm{p}-\tau_\mathrm{p0})+\TCL_{\mathrm{p0}} +\mathcal{O}(c^{-4})
\end{eqnarray}
with
\begin{equation}
\label{eq:LP}
	L_\mathrm{P} \equiv \frac{3}{2}\frac{GM_{\mathrm{M}}}{c^{2} \bar{a}_\mathrm{p}} \left[ 1 + \frac{7}{3} J^{\M}_2\frac{R^2_{\M}}{\bar{a}^2_{\mathrm{p}}} \left(1-\frac{3}{2}\sin^2 \bar{i}_{\mathrm{p}}\right) \right],
\end{equation}
where $\bar{a}_\mathrm{p}$, $\bar{e}_\mathrm{p}$ and $\bar{i}_{\mathrm{p}}$ are the mean semi-major axis, eccentricity and inclination of the orbit, respectively,
and we also neglect the higher-order spherical harmonic components of $U_\M$.

If we choose TCL for the coordinate simultaneity of these two clocks' proper times \eqref{eq:taus2TCL} and \eqref{eq:taup2TCL}, we can have 
\begin{equation}
    \tau_{\mathrm{s}} = \frac{1+L_\mathrm{P}}{1+\LL} \tau_{\mathrm{p}} + \tau_{\mathrm{s}0} - \frac{1+L_\mathrm{P}}{1+\LL}  \tau_{\mathrm{p}0} + \frac{\TCL_{\mathrm{p0}}-\TCL_{\mathrm{s0}}}{1+\LL}.
\end{equation}
After adjustment of initial constants in the above equation, we might align the orbital proper time $\tau_{\mathrm{p}}$ to the selenoid proper time $\tau_{\mathrm{s}}$, i.e. $\tau_{\mathrm{s}}=\tau_{\mathrm{p}}$  , as long as $L_\mathrm{P}=\LL$, leading to a specific mean semi-major axis that
\begin{equation}
\label{eq:apLLJ2}
   \bar{a}_\mathrm{p}=\frac{3}{2}\frac{GM_{\mathrm{M}}}{c^{2}\LL}\left[ 1 + \frac{28}{27} J_2^{\M} \LL^2 \left(\frac{GM_{\mathrm{M}}}{c^{2}R_{\M}}\right)^{-2} \left(1-\frac{3}{2}\sin^2 \bar{i}_{\mathrm{p}}\right)\right]
\end{equation}
where we neglect the nonlinear effect of $J_2^{\M}$. 
We call such an orbit as the ``\tao''. 
It has two distinctive properties that
(i) the proper time of an ideal clock in the \tao\ could naturally equal to the proper time of an ideal clock on the selenoid,
i.e. $\tau_{\mathrm{s}}=\tau_{\mathrm{p}}$;
and (ii) TCL could be easily found by scaling of the proper time of an ideal clock in the \tao\ with a known factor related the potential of the selenoid, i.e. $\TCL=(1+\LL)\tau_{\mathrm{p}}+\mathrm{constant}$.

When $J_2^{\M}$-term could be dropped in Eq.~\eqref{eq:apLLJ2} by picking up a specific inclination that $3\sin^2 \bar{i}_{\mathrm{p}}=2$,
we can obtain that the mean semi-major axis of the \tao\ around the Moon is
\begin{equation}
   \bar{a}_\mathrm{p}=\frac{3}{2}\frac{GM_{\mathrm{M}}}{c^{2}\LL}=2605.9\ \mathrm{km}
\end{equation}
where we adopt $\LL=3.14027\times10^{-11}$ \citep{Ardalan2014CMDA118.75}. 
In fact, by making use of Eqs.~\eqref{eq:taus2TCL}, \eqref{eq:LL}  and \eqref{eq:taup2TCL}, we can have that
\begin{equation}
   \bar{a}_\mathrm{p} \approx \frac{3}{2}R_{\M}\left[1+J_2^{\M}\left(\frac{29}{54}-\frac{14}{9}\sin^2\bar{i}_{\mathrm{p}}\right)-\frac{1}{2}\eta_{\M}\right]\approx 1.5\, R_{\M},
\end{equation}
since $J_2^{\M}\ll1$ and $\eta_{\M}\ll1$ for the Moon.
Therefore, we can find a more helpful insight that
if a nearly spherical body has its rotational surface speed much less than its first cosmic speed,
then it has a \tao\ above its surface with the height by about a half of the body's radius.

With the help of the International Astronomical Union Resolutions \citep{Soffel2003AJ126.2687,IAU2024ResII}, we might trace the proper time $\tau_\mathrm{p}$ of an ideal clock in the \tao\ back to UTC that 
\begin{equation}
 \UTC=(1-k)\,\tau_\mathrm{p} + P + \mathrm{const.},
\end{equation}
where the clock might have a significant frequency offset $k$ of about $6.5\times 10^{-10}=$ 56 $\muup$s d$^{-1}$ with respect to UTC, $P$ collects all of the periodic variations with a biggest amplitude of 1.6 ms, and the constant is a combination of the initial reading of the clock and other defining constants.

\section{Simulation}
\label{sec:sim}
In order to understand whether and to what extent a clock on the predicted \tao\ could realize the selenoid proper time in a more realistic lunar environment, we carry out a set of numerical simulations.
In these simulations, we include the point-mass gravitational effects of the Moon, Sun and all planets and higher-order spherical harmonics of the Moon.
Table~\ref{tab:dynmod} gives a comparison of the model we used for theoretical analysis in Sect.~\ref{sec:theory} and the one for our numerical simulations.

\begin{table}[t]
	\caption{Comparison of the models for theoretical analysis and numerical simulation\newline}
	\label{tab:dynmod}
	\centering
	\begin{tabular}{p{0.25\linewidth}p{0.15\linewidth}p{0.3\linewidth}}
		\hline\hline 
		& Theoretical analysis & Numerical\par simulation \\
		\hline 
		Point-mass
		&  Moon 
		& Moon\par Sun\par 8 planets\\
		\hline 
		Nonspherical Effects
		& $J_2$ term of the Moon 
		& up to the 100th-degree harmonics of the Moon\\ 
		\hline 
	\end{tabular}
\end{table}

We choose 4 different orbits with their initial inclinations of $i_{\mathrm{p,0}}=\{0,\ 25^{\circ},\ 54.74^{\circ},\ 85^{\circ}\}$ and initial semi-major axes $a_\mathrm{p,0}$ calculated based on Eq.~\eqref{eq:apLLJ2}.
We propagate the trajectory of each orbit for a year.
By using Eq.~\eqref{eq:tau2TCL}, we can calculate the proper time $\tau^*_{\mathrm{p}}$ on each simulated orbit from TCL.
Therefore, we might obtain the desynchronization 
\begin{equation}
\label{eq:desyn}
    \Delta^* = \tau^*_{\mathrm{p}}-\tau_{\mathrm{s}},
\end{equation}
indicating how well a clock on the \tao\ could realize the selenoid proper time, and we might derive its frequency offset $\Delta f^*$ to tell the drift rate between these two proper times
\begin{equation}
    \Delta f^* = \left< \frac{\mathrm{d}\Delta^*}{\mathrm{d}\tau^*_{\mathrm{p}}} \right>,
\end{equation}
where $\left<\cdot\right>$ means average over a long-term time span.

\begin{figure}[t!]
  \centering
  \includegraphics[width=\hsize]{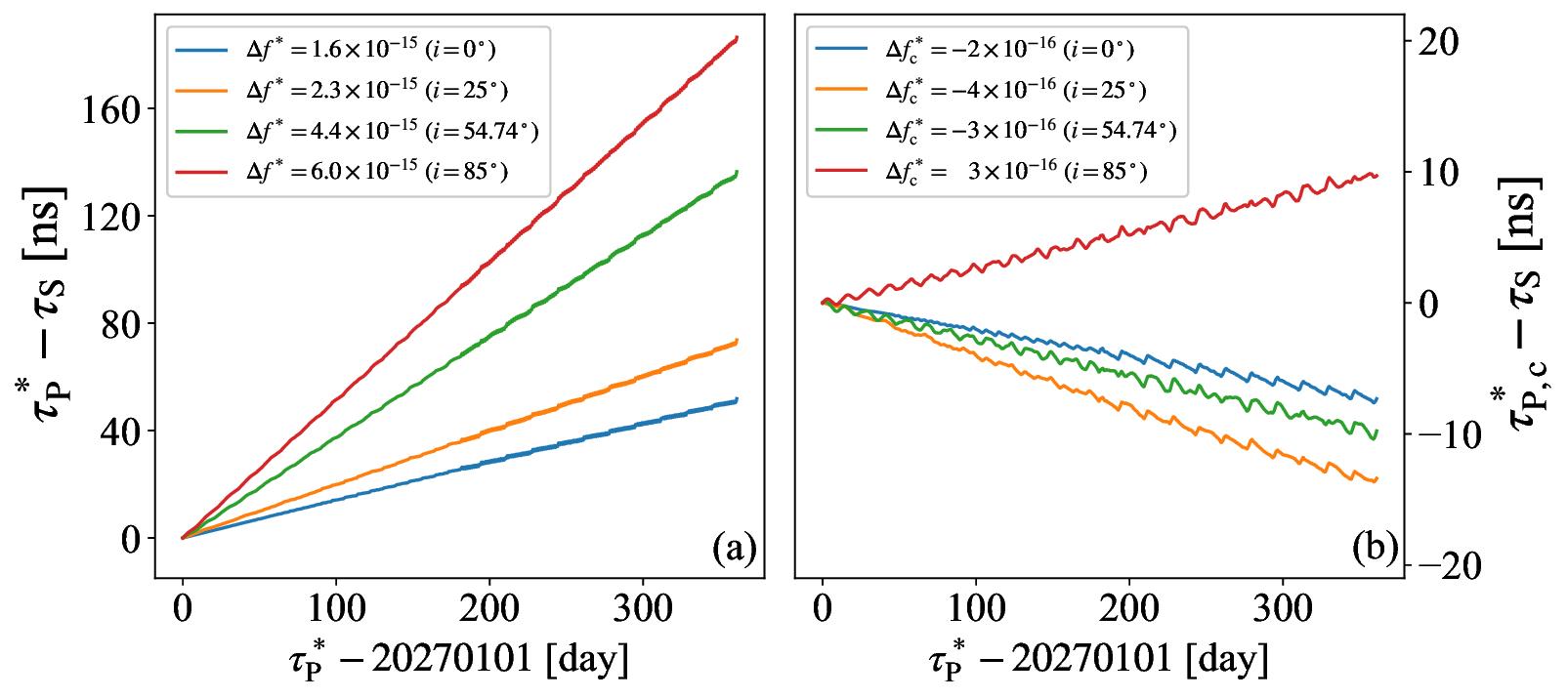}
   \caption{Left panel: the desynchronization $\Delta^*=\tau_{\mathrm{p}}^*-\tau_s$ and their frequency offsets $\Delta f^*$ of four numerically simulated \tao s with different initial inclinations. Right panel: the corrected $\Delta^*_{\mathrm{c}}=\tau^*_{\mathrm{p,c}}-\tau_s$ and $\Delta f^*_{\mathrm{c}}$ for the same four simulated orbits by accounting for the deviation of the mean orbital elements in  our simulations from the ones required by the \tao s.}
  \label{fig:dtau_t}
\end{figure}

Figure~\ref{fig:dtau_t}(a) shows that the desynchronization $\Delta^*$ grow with the increment of the initial inclination, from $\Delta^*=40$ ns for $i_{\mathrm{p,0}}=0$ to  $\Delta^*=190$ ns for $i_{\mathrm{p,0}}=85^\circ$ after a year. 
It suggests that the frequency offset $\Delta f^*$ is at the level of $\lesssim 6 \times 10^{-15}$ (see Table~\ref{tab:aeik} for details).
These numbers might demonstrate the deviations in the time and frequency domains of an ideal \tao\ from a more realistic one with more gravitational perturbations.
After a comparison with a uncertainty in the  realization of selenoid proper time O2, 
we find that the offset $\Delta f^*$  of the \tao\ is no more than 3.75\% of the frequency difference of $1.6\times10^{-13}$ in O2 due to the high variations of the lunar surface topography \citep{Bourgoin2025arXiv2507.21597}.
It might suggest that the realization of O2 by deploying clocks in the \tao\ would have less susceptibility to interference from natural causes than the one by landing clocks on the lunar surface.  

We hypothesize that the deviation of the mean orbital elements in our numerical simulations from those required  by the \tao~\eqref{eq:apLLJ2} might cause those desynchronization and frequency offset in Fig.~\ref{fig:dtau_t}(a).
In order to test this hypothesis, we correct the simulated $\tau^*_{\mathrm{p}}$, $\Delta^*$ and $\Delta f^*$ as
\begin{eqnarray}
    \tau^*_{\mathrm{p,c}} & = &  \tau^*_{\mathrm{p}} - \Delta L_\mathrm{P}\,\tau^*_{\mathrm{p}},\\
    \Delta^*_{\mathrm{c}} & = & \tau^*_{\mathrm{p,c}} - \tau_{\mathrm{s}},\\
    \Delta f^*_{\mathrm{c}} & = & \Delta f^* - \Delta L_\mathrm{P},
\end{eqnarray}
with
\begin{eqnarray}
\label{eq:DLP}
    \Delta L_\mathrm{P} & = & L_\mathrm{P}(\bar{a}^*_{\mathrm{p}},\bar{i}^*_{\mathrm{p}}) - L_\mathrm{P}(\bar{a}_{\mathrm{p}},\bar{i}_{\mathrm{p}}) \nonumber\\
    & = & - \frac{3}{2}\frac{GM_{\mathrm{M}}}{c^{2} \bar{a}_\mathrm{p}}  \frac{\Delta a}{\bar{a}_\mathrm{p}} \nonumber\\
    & & - \frac{21}{2} J^{\M}_2 \frac{GM_{\mathrm{M}}}{c^{2} \bar{a}_\mathrm{p}}  \frac{R^2_{\M}}{\bar{a}^2_{\mathrm{p}}} \left(1-\frac{3}{2}\sin^2 \bar{i}_{\mathrm{p}}\right) \frac{\Delta a}{\bar{a}_\mathrm{p}}\nonumber\\
    & & -\frac{21}{2} J^{\M}_2 \frac{GM_{\mathrm{M}}}{c^{2} \bar{a}_\mathrm{p}}  \frac{R^2_{\M}}{\bar{a}^2_{\mathrm{p}}} \sin \bar{i}_{\mathrm{p}} \cos \bar{i}_{\mathrm{p}}\, \Delta i\nonumber\\
    & & +\mathcal{O}[(\Delta a)^2,\ \Delta a\,\Delta i,\ (\Delta i)^2],
\end{eqnarray}
where $L_{\mathrm{P}}$ is defined in Eq.~\eqref{eq:LP},  $\bar{a}^*_{\mathrm{p}}$ and $\bar{i}^*_{\mathrm{p}}$ are the mean elements obtained by averaging outcomes of numerical simulations,
and we neglect the nonlinear effects of $\Delta a= \bar{a}^*_{\mathrm{p}}-\bar{a}_{\mathrm{p}}$ and $\Delta i = \bar{i}^*_{\mathrm{p}}-\bar{i}_{\mathrm{p}}$.
Since $J^{\M}_2\ll1$, we could expect the first term of $\Delta a$ in Eq.~\eqref{eq:DLP} plays the most important role there.
Figure~\ref{fig:dtau_t}(b) shows that the absolute values of corrected $\Delta^*_{\mathrm{c}}$ are no more than 13 ns  after a year and the absolute corrected $\Delta f^*_{\mathrm{c}}$ is no more than $4\times 10^{-16}$.
It might suggest that a more careful deployment of a clock into the \tao\ could improve its performance for realizing O2 by a factor of 10.

\begin{table*}
\renewcommand{\arraystretch}{1.2}
	\centering
	\caption{Comparison of nominal mean orbital elements $\bar{\sigma}_{\mathrm{p}}$ and mean orbital elements $\bar{\sigma}^*_{\mathrm{p}}$ from our numerical simulations with $\sigma=\{a,\ e,\ i\}$. The resulting and corrected frequency offsets are also listed.}
	\begin{tabular}{ccccccccc}
		\hline\hline\par
		Number &$\bar{a}_{\mathrm{p}}$ [km] & $\bar{e}_{\mathrm{p}}$ & $\bar{i}_{\mathrm{p}}$ & $\bar{a}^*_{\mathrm{p}}$ [km] & $\bar{e}^*_{\mathrm{p}}$ & $\bar{i}^*_{\mathrm{p}}$ & $\Delta f^*$ & $\Delta f^*_\mathrm{c}$\\
		\hline
		1&2606.2658&0&0&2606.1094&0.0039&$1.907^\circ$&$1.6\times10^{-15}$&$-2\times10^{-16}$\\
		2&2606.1186&0&$25^\circ$&2605.9063&0.0056&$23.395^\circ$&$2.3\times10^{-15}$&$-4\times10^{-16}$\\
		3&2605.7163&0&$54.736^\circ$&2605.3468&0.0046&$53.475^\circ$&$4.4\times10^{-15}$&$-3\times10^{-16}$\\
		4&2605.4477&0&$85^\circ$&2604.9803&0.0040&$84.715^\circ$&$6.0\times10^{-15}$&$\phantom{-}3\times10^{-16}$\\
		\hline
	\end{tabular}
	\tablefoot{In our simulations, we adopt each initial states as $\sigma_{\mathrm{p,0}}=\bar{\sigma}_{\mathrm{p}}$.  }
	\label{tab:aeik}
\end{table*}

\section{Conclusions and discussion}
\label{sec:con}
\begin{table}[t!]
	\caption{Mean semi-major axis $\bar{a}_{\mathrm{p}}$ of the \tao\ for 4 terrestrial planets\newline}
	\label{tab:TAO_a}
	\centering
	\begin{tabular}{lc}
		\hline\hline 
		Planet & $\bar{a}_{\mathrm{p}}$ [km] \\
		\hline 
		Mercury	& 3660.097\\
		Venus	& 9097.728\\ 
		Earth   & 9556.250\\
        Mars    & 5087.696\\
		\hline 
	\end{tabular}
	\tablefoot{In the above cases, the orbit inclination $\bar{i}_{\mathrm{p}}$ is set to be 0.  }
\end{table}

In the context of definition of the lunar reference time and facing the challenges for landing clocks on the surface of the Moon,
we show that there exist the \tao s around the Moon
with its semi-major axis of about 1.5 lunar radius.
The readings of an ideal clock in such an orbit might equal to the selenoid proper time
and the same readings might be easily converted to TCL by a known linear transformation. 
Therefore, it could be possible to simultaneously realize the lunar reference time options O1 and O2 of \citet{Bourgoin2025arXiv2507.21597} by a single orbital clock.
In order to assess its performance, we conduct a set of numerical simulations.
We find that the proper time in the \tao\ under a more realistically lunar gravitational environment would desynchronize from the selenoid proper time up to 190 ns after a year with a frequency offset of $6\times10^{-15}$, 
which is solely 3.75\% of the frequency difference in O2 caused by the lunar surface topography.
Meanwhile, if we could account for the deviation of the mean orbital elements in our simulations from the ones required by the \tao s, 
we would reduce the proper times' desynchronization and frequency offset by an order of magnitude to $13$ ns and $4\times 10^{-16}$.

Besides the Moon, the terrestrial planets might have their own \tao s as well (see Table~\ref{tab:TAO_a}). 
It means that it might be possible to realize the reference times of planets beyond the Earth-Moon system with clocks in these orbits, showing the scalability of the options built on the \tao s and reducing the risk of landing clocks on the surfaces of these planets.

\begin{acknowledgements}
      This work is funded by
the Strategic Priority Research Program on Space Science of the Chinese Academy of Sciences (XDA300103000, XDA30040000, XDA30030000, XDA0350300, XDA30040500 and XDA0350404) and
 the National Natural Science Foundation of China (Grants No. 62394350, No. 62394351 and No. 12273116).
 J.Z. is funded by the Jiangsu Funding Program for Excellent Postdoctoral Talent (Grant No. 2025ZB724).
\end{acknowledgements}

%

\bibliographystyle{aa}
\bibliography{refs}

@ARTICLE{Ardalan2014CMDA118.75,
       author = {{Ardalan}, A.~A. and {Karimi}, R.},
        title = "{Effect of topographic bias on geoid and reference ellipsoid of Venus, Mars, and the Moon}",
      journal = {\cmda},
         year = 2014,
        month = jan,
       volume = {118},
       number = {1},
        pages = {75-88},
          doi = {10.1007/s10569-013-9523-6}
}

@ARTICLE{Bourgoin2025arXiv2507.21597,
       author = {{Bourgoin}, Adrien and {Defraigne}, Pascale and {Meynadier}, Fr{\'e}d{\'e}ric},
        title = "{Lunar Reference Timescale}",
      journal = {arXiv e-prints},
         year = 2025,
        month = jul,
          doi = {10.48550/arXiv.2507.21597},
archivePrefix = {arXiv},
       eprint = {2507.21597},
 primaryClass = {gr-qc},
       adsurl = {https://ui.adsabs.harvard.edu/abs/2025arXiv250721597B},
      note = {\,Lunar Reference Timescale, \url{https://arxiv.org/abs/2507.21597}}
}

@ARTICLE{Formichella2021GPSSol25.56,
       author = {{Formichella}, Valerio and {Galleani}, Lorenzo and {Signorile}, Giovanna and {Sesia}, Ilaria},
        title = "{Time--frequency analysis of the Galileo satellite clocks: looking for the J2 relativistic effect and other periodic variations}",
      journal = {GPS Solut.},
         year = 2021,
        month = feb,
       volume = {25},
       number = {2},
        pages = {56},
          doi = {10.1007/s10291-021-01094-2}
}

@misc{IAU2024ResII,
  title        = {Resolution II: ``to establish a standard Lunar Celestial Reference System (LCRS) and Lunar Coordinate Time (TCL)''},
  author       = {IAU},
  year         = 2024,
  note         = {\url{https://www.iau.org/Iau/Publications/List-of-Resolutions}}
}

@misc{IAU2024ResIII,
  title        = {Resolution III: ``on the establishment of a coordinated lunar time standard by international agreement''},
  author       = {IAU},
  year         = 2024,
  note         = {\url{https://www.iau.org/Iau/Publications/List-of-Resolutions}}
}

@ARTICLE{Nelson2011Metrologia48.S171,
       author = {{Nelson}, Robert A.},
        title = "{Relativistic time transfer in the vicinity of the Earth and in the solar system}",
      journal = {Metrologia},
         year = 2011,
        month = aug,
       volume = {48},
       number = {4},
        pages = {S171-S180},
          doi = {10.1088/0026-1394/48/4/S07}
}

@ARTICLE{Kouba2004GPSSol8.170,
       author = {{Kouba}, J.},
        title = "{Improved relativistic transformations in GPS}",
      journal = {GPS Solut.},
         year = 2004,
        month = jul,
       volume = {8},
       number = {3},
        pages = {170-180},
          doi = {10.1007/s10291-004-0102-x}
}

@ARTICLE{Soffel2003AJ126.2687,
      author = {{Soffel}, M. and {Klioner}, S.~A. and {Petit}, G. and {Wolf}, P. and {Kopeikin}, S.~M. and {Bretagnon}, P. and {Brumberg}, V.~A. and {Capitaine}, N. and {Damour}, T. and {Fukushima}, T. and {Guinot}, B. and {Huang}, T. -Y. and {Lindegren}, L. and {Ma}, C. and {Nordtvedt}, K. and {Ries}, J.~C. and {Seidelmann}, P.~K. and {Vokrouhlick{\'y}}, D. and {Will}, C.~M. and {Xu}, C.},
        title = "{The IAU 2000 Resolutions for Astrometry, Celestial Mechanics, and Metrology in the Relativistic Framework: Explanatory Supplement}",
      journal = {\aj},
         year = 2003,
        month = dec,
      volume = {126},
      number = {6},
        pages = {2687-2706},
          doi = {10.1086/378162}
}

\begin{appendix}




\end{appendix}
\end{CJK*}
\end{document}